\documentclass[12pt]{article}
\usepackage{epsfig}

\def\beq{\begin{equation}}
\def\eeq#1{\label{#1}\end{equation}}
\def\eeqn{\end{equation}}

\def\beqa{\begin{eqnarray}}
\def\eeqa#1{\label{#1}\end{eqnarray}}
\def\eeqan{\end{eqnarray}}

\let\bar=\overbar

\def\Dslash{\not{\hbox{\kern-4pt $D$}}}
\def\dslash{\not{\hbox{\kern-2pt $\del$}}}

\def\msb{{\bar{\ssstyle M \kern -1pt S}}}

\usepackage{fancyhdr,graphicx}
\def\Title#1{\begin{center} {\Large {\bf #1} } \end{center}}

\begin{document}

\Title{Dark matter effect on the mass measurement of neutron stars}

\bigskip\bigskip


\begin{raggedright}

{\it Ang Li\index{Vader, D.}\\
Department of Astronomy and Institute of Theoretical Physics and Astrophysics\\
Xiamen University\\
422 Siming South Road\\
Xiamen 361005, Fujian province \\
China\\
{\tt Email: liang@xmu.edu.cn}}
\bigskip\bigskip
\end{raggedright}

\section{Introduction}

The recent measurement of the Shapiro delay in the radio pulsar PSR
J1614-2230 yielded a mass of 1.97 $\pm$ 0.04
$M_{\odot}$~\cite{Dem10}. Such a high neutron star (NS) mass might
rule out many predictions of non-nucleonic components (free quarks,
mesons, hyperons) in NS interiors, since they usually reduce the
theoretical maximum mass of the star
\cite{Bur10,Bur11,Zuo04,Li06,Li08a,Li08l,Peng08,Li09,Li10k}. For
example, a large NS maximum mass larger than 2 $M_{\odot}$ is
obtained from nucleonic equation of state (EoS) from the microscopic
Brueckner theory, but a rather low value below 1.4 $M_{\odot}$ is
found for hyperon stars (HSs) in the same method \cite{Bur10,Bur11}.
Although the present calculation did not include three-body hyperon
interaction due to the complete lack of experimental and theoretical
information, it seems difficult to imagine that these could strongly
increase the maximum mass, since the importance of the hyperon
interactions should be minor as long as the hyperonic partial
densities remain limited. However, if there is a universal strong
repulsion in all relevant channels the maximum mass may be
significantly raised~\cite{Vid11}, so the including of the hyperonic
three-body interaction, together with an improved hyperon-nucleon
and hyperon-hyperon potentials, is still appealing to settle this
apparent contradiction. In addition, the presence of a
strongly-interacting quark matter in the star's interior, i.e., a
hybrid star model, is proposed to be a good candidate for
troubleshooting this problem \cite{Sch11}, However, NS masses
substantially above 2 $M_{\odot}$ seem to be out of reach even for
hybrid stars using most of effective quark matter EoS (Eg.~bag
model~\cite{Bur02}, Nambu-Jona-Lasinio model~\cite{Bal03}, color
dielectric model~\cite{Mai04}). A hybrid star with 2 $M_{\odot}$ is
only allowed for the description of quark matter in the
Dyson-Schwinger approach~\cite{Che11}, or possible in the latest
version of the Polyakov-Nambu-Jona-Lasinio model~\cite{david11}.

Very recently, dark matter~(DM), as another possible constituent in
NS interior, has been taken into account and a new type of compact
star, i.e., DM-admixed NS, has been studied in several articles
\cite{Gar10,Leu11,San11,Gol11,San09,Ber08,Kou08,Kou10,Lav10}. The
general effect induced by DM inside NS is complicated due to the
lack of information about the particle nature of DM. DM could
annihilate, such as the most favored candidate, neutralino, which
may lead to sizable energy deposit and then enhance the thermal
conductivity or trigger the deconfinement phase transition in the
core of NS for the emergency of a quark star, as illustrated by
Perez-Garcia et al. in \cite{Gar10}. Such quark star objects are at
present very uncertain in theory and could easily accord with
astrophysical measurements within the modification of model
parameters \cite{Li10,Li11}. Another generally considered DM
candidate is the non-self-annihilating particle, such as the newly
interesting mirror DM (\cite{mir} and references therein) or
asymmetric DM (\cite{asy} and references therein). When they
accumulate in NSs, the resulting maximum mass is then rather
sensitive to the EoS model of DM. Assuming that the DM component is
governed by an ideal Fermi gas, Leung et al. \cite{Leu11} studied
the various structures of the DM-admixed NSs by solving the
relativistic two-fluid formalism. Ciarcelluti \& Sandin \cite{San11}
approximated the EoS of mirror matter with that of ordinary nuclear
matter, varied the relative size of the DM core, and explained all
astrophysical mass measurements based on one nuclear matter EoS.

In this paper, we will consider non-self-annihilating DM particles
as fermions, and the repulsive interaction strength among the DM
particles is assumed to be a free parameter $m_\mathcal{I}$ as in
\cite{Nar06}. Different to previous DM-admixed NS models,  we take
the total pressure (energy) density as the simple sum of the DM
pressure(energy) and NS pressure (energy), the general dependence of
the mass limit on DM particle mass and the interaction strength is
then presented based on the present model. Furthermore, in the
particular case of PSR J1614-2230, we notice that the DM around the
star should also contribute to its mass measurement due to the pure
gravitational effect. However, our numerically calculation
illustrates that such contribution could be safely ignored because
of the usual diluted DM environment assumed.

The paper is arranged as follows. The details of our theoretical
model are presented in \S2, followed by the numerical results. A
short summary is given in \S3.

\section{The model}
\label{sect:model}

We begin with the treatment of the DM particles scattered inside the
star. They would modify the local pressure-energy density
relationship of the matter and hence change the theoretical
prediction of the gravitational mass of the star. The structure
equations for compact stars, the Tolman-Oppenheimer-Volkov equations
are written as:
\begin{equation}
 \frac{dP(r)}{dr}=-\frac{Gm(r)\mathcal{E}(r)}{r^{2}}
 \frac{\Big[1+\frac{P(r)}{\mathcal{E}(r)}\Big]
 \Big[1+\frac{4\pi r^{3}P(r)}{m(r)}\Big]}
 {1-\frac{2Gm(r)}{r}},
   \label{tov1:eps}
\end{equation}
\begin{equation}
\frac{dm(r)}{dr}=4\pi r^{2}\mathcal{E}(r),
  \label{tov2}
\end{equation}
being $G$ the gravitational constant. $P$ and $\mathcal{E}$ denote
the pressure and energy density. The EoS of the star, relating $P$
and $\mathcal{E}$, is needed to solve the above set of equations. In
our DM-admixed NS model, $P = P_N + P_{\chi}, \mathcal{E} =
\mathcal{E}_N +\mathcal{E}_{\chi}$, with the subscript $N(\chi)$
representing NS matter (DM).

The EoS of the ordinary NS matter is handled in the following way:
(i) We treat the interior of the stars as $\beta$-equilibrium
hypernuclear matter, with certain amount of leptons to maintain
charge neutrality. The hadronic energy density we use in the article
is based on the microscopic parameter-free Brueckner-Hartree-Fock
nuclear many-body approach, employing the latest derivation of
nucleon-nucleon microscopic three-body force \cite{Li08},
supplemented by the very recent Nijmegen extended soft-core ESC08b
hyperon-nucleon potentials \cite{Sch11}. The EoS can be computed
straightforwardly after adding the contributions of the
noninteracting leptons \cite{Sch11}. (ii) For the description of the
NS crust, we join the hadronic EoS with those by Negele and
Vautherin~\cite{nv} in the medium-density regime, and those by
Feynman-Metropolis-Teller  \cite{fmt} and
Baym-Pethick-Sutherland~\cite{bps} for the outer crust.

The EoS of DM is modeled to be that of a self-interacting Fermi gas
with one parameter $m_\mathcal{I}$ accounting for the energy scale
of the interaction \cite{Nar06}. For weak interaction (WI) the scale
$m_\mathcal{I}$ can be interpreted as the expected masses of W or Z
bosons generated by the Higgs field, which is $\sim300$ GeV. For
strongly interacting (SI) DM particles, $m_\mathcal{I}$ is assumed
to be $\sim100$ MeV, according to the gauge theory of the strong
interactions. This is a wide enough range of energy scale, and we
hope the calculation would cover most of the promising DM
candidates.
\begin{figure}[htb]
\begin{center}
\epsfig{file=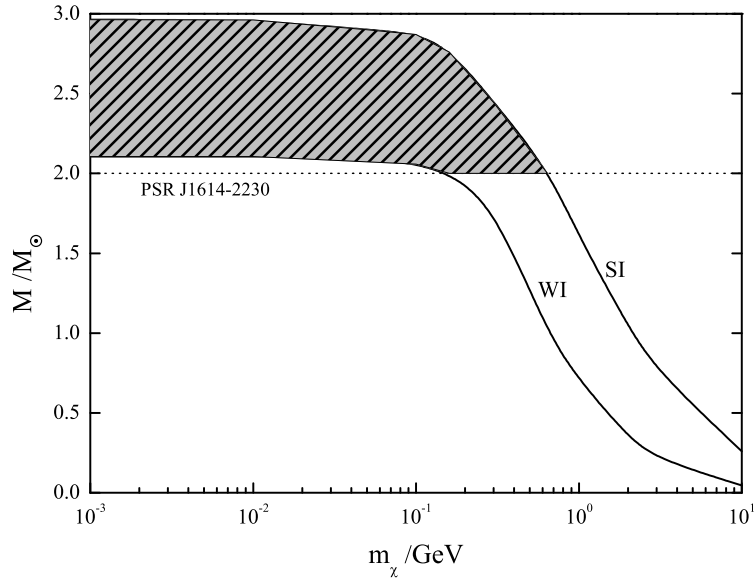,height=3.5in} \caption{HSs' maximum masses as
a function of the particle mass of DM candidates $m_{\chi}$.
 The upper line is for $m_{\mathcal{I}}$ = 100 MeV (SI case),
 and the lower line for $m_{\mathcal{I}}$ = 300 GeV (WI case).
 Again the $\sim 2$ $M_{\odot}$ limit of PSR J1614-2230 is indicated
 with a horizontal line. Taken from Ref.~\cite{Li12}.} \label{fig1}
\end{center}
\end{figure}

Then we consider the mass contribution of DM halo via gravitational
capture. We first estimate the size of the halo $R$ as big as that
of the possible Roche lobe~\cite{Egg} of the centered PSR
J1614-2230, namely $ R = 0.49 (M_1/M_2)^{2/3}~a/ \{0.6
(M_1/M_2)^{2/3} + ln[1+(M_1/M_2)^{1/3}]\} $, with $a$ being the
major semi-axis of this binary system, namely 3$\times$10$^{11}$ cm.
$M_1$($M_2$) is the gravitational mass of the NS (the companion
star). The local DM density is estimated from several spherically
symmetric Galactic DM profiles~\cite{Spr08,Nav97,Bur95,Sal00}, and
we take an average value of $\bar{\rho_{\chi}}$ = 0.474 GeV/cm$^3$.
Thus the contributed mass of gravitationally captured DM particles
can be obtained from $M_{\chi}= \frac{4}{3}\pi R^3
\bar{\rho_{\chi}}$. Here the size of the star ($\sim10$ km) has been
neglected compared to its large Roche lobe ($\sim10^6$ km), and the
halo has been regarded as an ideal spherical object.

Finally our calculated result is present in Fig.~\ref{fig1} with a
shallowed area indicating the mass limit of DM particle $m_{\chi}$,
assuming the observed PSR J1614-2230 is a HS. The $\sim 2$
$M_{\odot}$ limit is indicated with a horizontal line. The upper
line corresponds to $m_{\mathcal{I}}$ = 100 MeV (SI case), and the
lower line to $m_{\mathcal{I}}$ = 300 GeV (WI case). Specially we
mention that the extra mass measurement contribution from the
extended halo is only around 10$^{-24}$ $M_{\odot}$, which could be
safely ignored for the analysis of Shapiro delay. This mass limit
from the compact star can be referred as an upper limit for the mass
of non-self-annihilating DM particles, namely, it should obey
$m_{\chi} < 0.64$ GeV for SI DM and $ < 0.16$ GeV for WI DM.

\section{Summary}

In this article, we consider
DM as another possible constituent in HSs'
interior, to solve the recent hyperon puzzle provoked by the
recent 2-solar-mass NS measurement. We take DM as
self-interacting Fermi gas with certain repulsive interaction among
the DM particles and non-interaction between DM
and ordinary matter as is generally assumed. We find that the star
maximum mass is sensitive to the particle mass of DM, and a
high enough star mass larger than 2 $M_{\odot}$ could be achieved
when the particle mass is small enough. In this particular model, a
strong upper limit $0.64$ GeV for DM mass is obtained in
SI DM and $0.16$ GeV for DM mass in WI DM. 
In order to relax such strong constraint, we further consider
the possible extended DM halo contribution to the particular mass
measurement. However, due to the diluted DM environment, such kind
of contribution could be safely ignored.

\bigskip
This work is funded by the National Basic Research Program of China
(Grant No 2009CB824800), the National Natural Science Foundation of
China (Grant No 10905048, 10973002, 10935001, 11078015), and the
John Templeton Foundation.

\end{document}